\documentclass[prl,twocolumn,letterpaper,showpacs]{revtex4}

\usepackage{graphicx}

\begin{document}

\title{Wave localization in strongly nonlinear Hertzian chains with mass defect}
\author{St\'ephane Job$^1$, Francisco Santibanez$^2$, Franco Tapia$^2$, and Francisco Melo$^2$}
\affiliation{$^{1}$Supmeca, 3 rue Fernand Hainaut, 93407 Saint-Ouen Cedex, France.
$^{2}$Departamento de F\'{\i}sica and Center for Advanced Interdisciplinary Research in Materials (CIMAT), Universidad de Santiago de Chile, Avenida Ecuador 3493, Casilla 307, Correo 2, Santiago de Chile.}

\date{\today}


\begin{abstract}
We investigate the dynamical response of a mass defect in a one-dimensional non-loaded horizontal chain of identical spheres which interact via the nonlinear Hertz potential. Our experiments show that the interaction of a solitary wave with a light intruder excites localized mode. In agreement with dimensional analysis, we find that the frequency of localized oscillations exceeds the incident wave frequency spectrum and nonlinearly depends on the size of the intruder and on the incident wave strength. The absence of tensile stress between grains allows some gaps to open, which in turn induce a significant enhancement of the oscillations amplitude. We performed numerical simulations that precisely describe our observations without any adjusting parameters.
\end{abstract}


\maketitle


Wave localization has been studied for long in linear systems~\cite{Lifshitz43} or lattices~\cite{SolidState55} and became widely studied more recently in nonlinear systems~\cite{Vakakis96}. For example, the presence of an isotope in a perfect linear crystal is known to enhance optical waves absorption at given frequencies~\cite{Sievers64}. One-dimensional chains of beads interacting via the Hertz potential are systems suitable to observe nonlinear localization effects. A loaded chain of identical beads is dispersive, allowing small perturbations to propagate as linear or weakly nonlinear acoustic waves~\cite{Nesterenko2001}. In contrast, when grains in a chain barely touch one another, the energy of an impulse only propagates as fully nonlinear solitary waves~\cite{Nesterenko2001,Coste1997,Job2005} resulting from the balance between dispersion and nonlinearity of the medium. Nesterenko early described this regime as a sonic vacuum limit~\cite{Nesterenko_1984_1985}. Dissipative effects such as viscoelasticity or friction only attenuate and spread these solitary waves~\cite{Poschel2001,Job2005}. In contrast, any heterogeneity of the medium capable of unbalancing dispersion and nonlinearity results in breaking the solitary wave symmetry. For example, a narrow pulse propagating in a chain of beads with decreasing sizes develops a long tail which spreads in time the momentum transfer~\cite{Doney2005,Melo2006,Job2007}. Designing powerfull impact protection systems takes advantage of these features~\cite{Hong05,Doney2005,Melo2006}. Granular chains made of successions of heavy and light beads also proved valuable efficiency in energy absorption~\cite{Doney2006}. More recently, fully nonlinear waves with finite-width were observed in chains containing periodic mass defects or soft inclusions~\cite{Porter2008}. Such nonlinear dimer chains are expected to support additionnal optical modes and forbidden band gap when subjected to a static load~\cite{Porter2008}.

The elementary interaction of either lighter or heavier intruders with solitary waves in non-loaded monodisperse chains of beads has been investigated numerically~\cite{Hascoet2000}. When a solitary wave reaches a mass defect, energy is partially reflected into a backward traveling solitary wave and is partially transmitted to the intruder. A heavy impurity slowly translates, leading to a large transmitted solitary waves train in the forward direction~\cite{Hascoet2000}, similarly to what was observed in stepped chains~\cite{Nesterenko_1984_1985,Nesterenko2001,Job2007}. A light intruder oscillates and scatters forward and backward weak delayed solitary waves trains~\cite{Hascoet2000}.

\begin{figure}[b]
\includegraphics{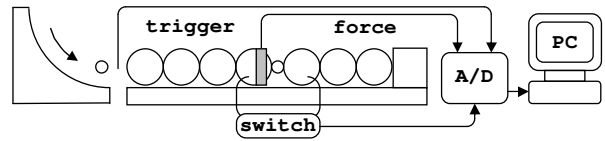}
\caption{\label{fig:fig01} Upper panel: Experimental setup showing a chain of beads with an intruder, sensors and acquisition facilities.}
\end{figure}

In this letter, we investigate experimentally the interaction of a solitary wave with a light intruder in a non-loaded monodisperse chain of beads. We observe that the impurity oscillates while interacting with a solitary wave. Oscillations remain localized and fade as soon as the wave travels away. The presence of a small distance gap between the intruder and its nearest neighbors enhances the oscillation amplitude. A multiscale analysis predicts the oscillations frequency as a function of intruder size and wave strength. High resolution numerical simulations capture all the experimental features.


Beads are {\em 100C6} steel roll bearings~\cite{Bead_Refs} with density $\rho=7780$~kg/m$^3$, Young's modulus $Y=203\pm4$~GPa and Poisson ratio $\nu=0.3$. The chain is made of $20$ equal beads (radius $R=13$~mm) and contains an intruder ($2.5\mbox{~mm}\leq R_i\leq10\mbox{~mm}$) in the middle, as shown in Fig.~\ref{fig:fig01}. The beads, barely touching one another, are aligned on a horizontal Plexiglas track. A special short track automatically aligns the intruder on the axis of the chain~\cite{Melo2006}. The chain is ended by a flat, fixed and heavy piece of steel. A nonlinear compressive wave (see below) is initiated from the impact of a small striker ($R_s=4$~mm). The pulse is monitored by measuring the load with a piezoelectric transducer ({\em PCB 200B02}, sensitivity $11.24$~mV/N and stiffness $1.9$~kN/$\mu$m) inserted inside a bead cut in two parts. The total mass of the active bead matches the mass of an original bead and the rigidity of the sensor is comparable to bead's material properties. The embedded sensor thus allows non-intrusive measurements of the force inside the chain~\cite{Job2005}.

\begin{figure}[b]
\includegraphics{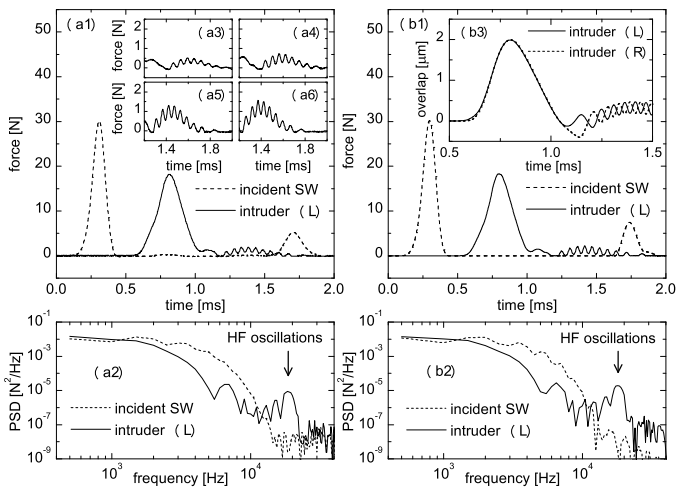}
\caption{\label{fig:fig02} Insets (a1-a6) are experiments and (b1-b3) are numerical simulations, both performed in a monodisperse chain containing a $3$~mm in radius intruder. Dashed line in (a1,b1) is the $6$ beads far the intruder incident solitary wave force versus time, and solid line is the force versus time at the left intruder's contact. Power spectral densities of these forces are plotted in (a2,b2), respectively, showing high frequency content in the intruder force. Closer views of oscillating tail in the intruder's force are shown in (a3-a6) for increasing incident force magnitude ($14.1$, $17.3$, $22.7$ and $26.3$~N, respectively). Solid and dashed lines in (b3) represent overlaps between the intruder and left and right neighbor beads, respectively.}
\end{figure}


Typical force signals measured at the intruder contact, plotted in Fig.~\ref{fig:fig02}a1, shows that large and well defined oscillations appear in the tail of the incident solitary wave. The tail corresponds to a slight reflection of the incident pulse on the intruder mass defect. Measured oscillations in the tail of the force are displayed in more details in Figs.~\ref{fig:fig02}a3-\ref{fig:fig02}a6 for increasing amplitudes of the incident solitary wave. Dissipative processes being negligible~\cite{Job2005}, conservative numerical simulations (see below) shown in Fig.~\ref{fig:fig02}b1 demonstrate satisfactory agreements with experiments, without any adjusting parameter. Calculated overlaps between the intruder and neighbor beads, shown in Fig.~\ref{fig:fig02}b3 (see also the relative displacements in Fig.~\ref{fig:fig04}a), demonstrate that force oscillations correspond to localized oscillations of the intruder. Figs.~\ref{fig:fig02}a2 and~\ref{fig:fig02}b2 depict power spectral densities of experimental and calculated forces, respectively, in which the high frequency component corresponds to the observed oscillations.

\begin{figure}[t]
\includegraphics{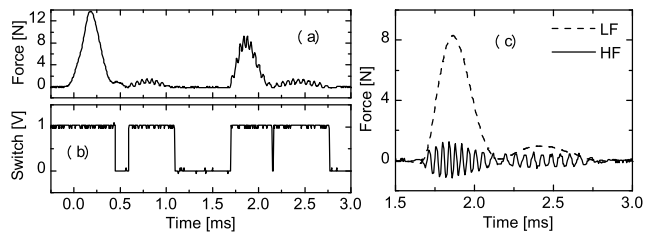}
\caption{\label{fig:fig03} Force versus time as detected by the embedded sensor in contact with a $3$~mm in radius intruder. The first large peak in (a) is the incident solitary wave, and the second one is the wave after being reflected by the rigid end. The gap opening between intruder and nearest beads is indicated in (b) by jumps from upper to lower level in the contact switch. High and low frequency components of the force, shown in (c), are obtained by using a tenth order Butterworth filter with zero phase distortion.}
\end{figure}

Fig.~\ref{fig:fig03} presents a longer force acquisition, in which the incident pulse and the pulse reflected by the rigid end are shown. Force signals exhibit features at two different time scales, that are separated by filtering low or high frequency contents, as shown in Fig.~\ref{fig:fig03}c. Oscillations are only observed when the intruder is loaded by the solitary wave. In Fig.~\ref{fig:fig03}a, intruder oscillations are only visible in the tail of the incident pulse ($0.0~\mbox{ms}<t<1.0~\mbox{ms}$). Stronger and relatively faster oscillations can be observed in the reflected pulse ($1.5~\mbox{ms}<t<3.0~\mbox{ms}$), both during the main compression and in the tail. Gap openings between the intruder and neighbor beads might explain the differences between incident and reflected pulses features. We designed an electrical switch (smooth metallic brushes in contact with neighbor beads, connected to a fast TTL electrical circuit, see Fig.~\ref{fig:fig01}) to detect any loss of contact. The switch is on $(1)$ when beads are in contact and off $(0)$ when loss of contact occurs. As shown in Fig.~\ref{fig:fig03}b, gap first opens right after the incident compression ($t\simeq0.5$~ms and $t\simeq2.1$~ms) and second at the end of tail ($t\simeq1.1$~ms and $t\simeq2.7$~ms). The first opening is due to the weak reflection of the pulse on the mass defect and the second one is due to the momentum and energy transfer from the oscillating intruder to neighbors. Higher amplitude oscillations appears provided a gap between the intruder and neighbor exists.

These observations are corroborated by numerical simulation. Calculated displacements of the intruder and neighbors shown in Fig.~\ref{fig:fig04}a and the overall gap around the intruder (the difference between left and right neighbor beads displacements) shown in Fig.~\ref{fig:fig04}b demonstrate that gaps open twice at the intruder, consistently with our experiments. Fig.~\ref{fig:fig04}c, showing the intruder displacement relative to the center of mass of the two neighbors beads, reveals that intruder oscillations even exist during the main compression of the incident pulse ($0.5~\mbox{ms}<t<1.0~\mbox{ms}$, see magnified view in Fig.~\ref{fig:fig04}d). Momentum transfer is enhanced by larger strain gradient in the presence of gaps.

\begin{figure}[t]
\includegraphics{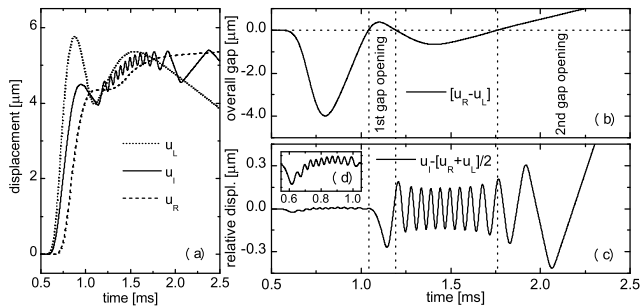}
\caption{\label{fig:fig04} Numerical simulations of the incident pulse, under same conditions as in Fig.~\ref{fig:fig02}b1. (a) Positions versus time of the intruder and left and right neighbor beads. (b) Overall gap around the intruder demonstrating gap openings occur when a solitary wave crosses the intruder. (c) Intruder displacement relative to the center of mass of the two neighbor beads. (d) Magnified relative displacement during the compression.}
\end{figure}

\begin{figure}[t]
\includegraphics{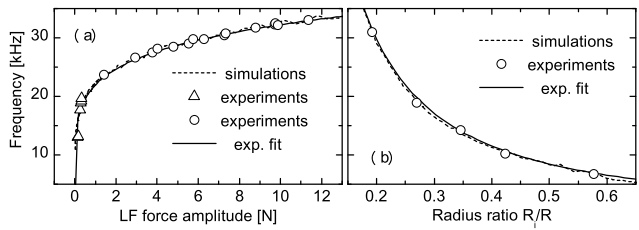}
\caption{\label{fig:fig05} Inset (a) shows frequency versus force's envelope amplitude, for an $R_{i}=2.5$~mm in radius intruder. Triangles and circles correspond to experimental frequency detected in the tail of the incident pulse and in the reflected pulse, respectively. Inset (b) shows frequency versus intruder's radius measured in the reflected signal, whose envelope amplitude was fixed to $\overline{F}_{m}=8.7\pm0.6$~N. Dashed lines correspond to numerical simulations and straight lines corresponds to Eq.~\ref{Eq:freq_vs_force_and_radius}.}
\end{figure}

Next, we analyzes how oscillations depend on wave strength and on intruder size. Frequency of oscillations is obtained from the analysis of power spectral densities of the force. Following measurements are repeated nine times and averaged to minimize errors. We first run a set of experiments at constant intruder size ($R_{i}=2.5$~mm) while varying the amplitude of the incident solitary wave. The force amplitude is obtained from the low-pass filtered signals and results are presented in Fig.~\ref{fig:fig05}a. We then keep the incident pulse amplitude constant ($8.6\pm0.5$~N) and test several intruder sizes, as depicted in Fig.~\ref{fig:fig05}b. Experiments show that the frequency of the intruder nonlinearly increases with the incident solitary wave strength and depends on the intruder size.


The physical behavior of solitary waves in chains of equal beads and implications for the existence of localized modes is summarized here. Under elastic deformation, the energy stored at the contact between two elastic bodies submitted to an axial compression corresponds to Hertz potential, $U_{H}=(2/5)\kappa\delta^{5/2}$, where $\delta$ is the overlap deformation between bodies, $\kappa^{-1}=(\theta+\theta')(R^{-1}+R'^{-1})^{1/2}$ and $\theta=3(1-\nu^2)/(4Y)$ are constants, and where $R$ and $R'$ are respective radii of curvature at the contact. The force felt at the interface derives from Hertz potential, $F_{H}=\partial_\delta U_H=\kappa\delta^{3/2}_{+}$. Index $+$ indicates that Hertz force is zero when the beads loose contact (no tensile force). The dynamics of a chain of beads is thus described by the following system of $N$ coupled nonlinear equations,
\begin{equation}\label{Eq:DiscreteEquation}
m\ddot{u}_n=\kappa\left[(u_{n-1}-u_{n})^{3/2}_{+}-(u_{n}-u_{n+1})^{3/2}_{+}\right],
\end{equation}
where $m$ and $u_n$ are the mass and the position of bead $n$, respectively. Considering long-wavelength perturbations, such that the strain $\psi=(-\partial_xu)\simeq(\delta/2R)\simeq(u/\lambda)\ll1$, a continuous equation can be derived from Eq.~\ref{Eq:DiscreteEquation}, which admits an exact solution~\cite{Nesterenko_1984_1985} in the form of a purely compressive and periodic traveling wave, $\psi(x,t)=\psi_m\cos^4[(x-vt)/(R\sqrt{10})]$. Wave speed, $v\propto\psi_m^{1/4}$, nonlinearly depends on maximum strain. Infinitely small perturbations in the acoustic limit propagate at zero speed, acoustic modes are thus forbidden. One hump of this periodic function represents a solitary wave solution~\cite{Nesterenko_1984_1985}. In addition to analytical estimations, we solve the nonlinear system of Eqs.~\ref{Eq:DiscreteEquation} by using a fourth order Runge-Kutta numerical scheme~\cite{Chatterjee1999}, the embedded force sensor being incorporated in simulations for closer comparisons~\cite{Job2008}. Numerical time step is few orders of magnitude smaller than the shortest physical duration and energy conservation is fulfilled within a relative error better than $10^{-9}$.


The characteristic frequency of localized oscillations can be estimated through a multiscale analysis of Eq.~\ref{Eq:DiscreteEquation}. Here, index $n$ denotes the intruder with radius $R_{i}$ and mass $m_{i}=(4/3)\pi\rho R_{i}^3$, and $\kappa_{i}=\kappa(R,R_{i})=\kappa(R_{i},R)$ is the elastic constant that depends on radii and properties of the beads. Index $n\pm1$ denote the two neighbor beads with radius $R>R_{i}$. We consider two distinct time scales: a slow variation of the order of solitary wave duration, and a fast time scale associated with intruder oscillations. Displacements of the intruder and neighbor beads are written as $u_{n}\simeq\overline{u}_{n}+\widetilde{u}_{n}\mbox{e}^{i\omega t}$ and $u_{n\pm1}\simeq\overline{u}_{n\pm1}$, where $\overline{u}_{n}$ and $\widetilde{u}_{n}$ are slowly varying functions of time. Harmonic oscillations amplitude is assumed negligible compared to solitary wave amplitude, $\widetilde{u}_{n}\ll\overline{u}_{n}$. Replacements into Eq.~\ref{Eq:DiscreteEquation} leads to,
\begin{eqnarray}
\ddot{\overline{u}}_{n} &\simeq& \frac{\kappa_{i}}{m_{i}}
\left[(\overline{u}_{n-1}-\overline{u}_{n})_{+}^{3/2}-(\overline{u}_{n}-\overline{u}_{n+1})_{+}^{3/2}\right],\\
\omega^2 &\simeq& \frac{3}{2}\frac{\kappa_{i}}{m_{i}}
\left[(\overline{u}_{n-1}-\overline{u}_{n})_{+}^{1/2}+(\overline{u}_{n}-\overline{u}_{n+1})_{+}^{1/2}\right],
\end{eqnarray}
where the first equation, at leading order, provides the displacement of the intruder at slow time scale. The second equation, at next order, determines the angular frequency of the oscillating intruder. We then introduce slowly varying forces at the contacts of the intruder, $\overline{F}_{-}=\kappa_i(\overline{u}_{n-1}-\overline{u}_{n})_{+}^{3/2}$ and $\overline{F}_{+}=\kappa_i(\overline{u}_{n}-\overline{u}_{n+1})_{+}^{3/2}$. Noticing that they almost behave in phase ($\overline{F}_{+}\simeq\overline{F}_{-}$, see Fig.~\ref{fig:fig02}b3), the first equation indicates that intruder acceleration oscillates around equilibrium position, $\ddot{\overline{u}}_{n}\simeq0$. The second equation provides the maximum oscillation frequency $f_m=\max{[\omega/2\pi]}$ as a function of the amplitude of the slow time force at the intruder contact, $\overline{F}_m\simeq\max{[\overline{F}_{+}]}\simeq\max{[\overline{F}_{-}]}$:
\begin{equation}\label{Eq:freq_vs_force_and_radius}
f_m \simeq \frac{3^{1/2}}{2\pi}\frac{\kappa_{i}^{1/3}\overline{F}_m^{1/6}}{m_{i}^{1/2}}
\simeq C\frac{(R/R_{i})^{4/3}\overline{F}_m^{1/6}}{(1+R_{i}/R)^{1/6}}
\end{equation}
where $C=(3/4\pi\sqrt{\pi\rho})/(2 \theta R^4)^{1/3}$. Eq.~\ref{Eq:freq_vs_force_and_radius} shows that oscillation frequency tends to zero when the load vanishes: oscillations stops as soon as the solitary wave leaves the intruder. Using the characteristics of our beads, we find a theoretical estimation, $C_{t}=2640$~Hz/N$^{1/6}$. Matching Eq.~\ref{Eq:freq_vs_force_and_radius} to experiments and simulations shown in Fig.~\ref{fig:fig05}, we find $C_{e}=2510\pm151$~Hz/N$^{1/6}$ and $C_{n}=2531\pm52$~Hz/N$^{1/6}$, respectively. The agreement between experiments, simulations and theory shown in Fig.~\ref{fig:fig05} is satisfactory, considering that no adjustable parameters is used between experiments and simulations.

It should be pointed out that weak delayed solitary wave trains, responsible for nonlinear leak of oscillating energy from the intruder to surrounding~\cite{Hascoet2000}, were detected in long duration acquisitions of the force recorded few beads before or after the intruder (not shown in this letter). This attenuation mechanism should lead to an exponential decay of the oscillating amplitude within a characteristic time much longer than the duration of the low frequency force envelope, $\tau_{leak}/\tau_{LF}\sim\sqrt{m/m_i}\gg1$, provided the mass of the intruder is lighter than a bead of the chain, $m_i\ll m$.

Eq.~\ref{Eq:freq_vs_force_and_radius} can also be obtained by considering the low frequency force envelope amplitude, of the order of $\overline{F}_m$, as a static load at the time scale of fast oscillations. The dynamics of a loaded monodisperse chain of beads of mass $m$ behaves according to linearized Hertz potential around static equilibrium; the intergrain stiffness reads $k\simeq(3/2)\kappa^{2/3}\overline{F}_m^{1/3}$. Small perturbations propagate as acoustic waves according to the dispersion relation $\omega=(2\sqrt{k/m})|\sin{(qR)}|$. Forcing a single particle to move at an angular frequency $\omega$ above the cutoff $\omega_c=2\sqrt{k/m}$ generates an evanescent wave since the wave number is $qR=\pi/2-j\cosh^{-1}{(\omega/\omega_c)}$. The group velocity $v_g=(\partial\Re{[q]}/\partial\omega)^{-1}$ is zero and the perturbation remains localized within a characteristic distance $x_c=(\Im{[q]})^{-1}=R/\cosh^{-1}{(\omega/\omega_c)}$. Localization achieves when replacing a single particle by a light intruder with mass $m_{i}\ll m$. Roughly estimating the angular frequency of the intruder from the free oscillation frequency between two heavy non-oscillating neighbors, $\omega_i\simeq\sqrt{2k_{i}/m_{i}}$ where $k_i\simeq(3/2)\kappa_i^{2/3}\overline{F}_m^{1/3}$, leads to Eq.~\ref{Eq:freq_vs_force_and_radius} expression. The frequency of oscillations exceeds the cutoff, $\omega_i\gg\omega_c$ and the energy remains localized within a characteristic distance that does not depend on static load and which is smaller than a single radius, $x_c\ll R$.


In conclusion, we have reported observations of energy localization in strongly nonlinear chains of elastic beads. Localization occurs when a light intruder oscillates at a time scale much shorter than the duration of the incident solitary wave. Oscillations are excited by the local strain gradient and enhanced by the presence of gap at the intruder contact. Oscillations are unable to radiate linear acoustic waves as they are forbidden in the sonic vacuum limit. Such nonlinear localization effects are likely to appear in polydisperse 3D granular assemblies. For instance, a significant number of weakly loaded light particles might play an important role on sound trapping at high frequency.


\begin{acknowledgments}
This work was supported by Conicyt under Fondap research program $N^{o}$ 11980002.
\end{acknowledgments}


\bibliography{loc}


\end{document}